\documentclass[conference, 10pt]{IEEEtran}

\usepackage{cite}
\usepackage{amsmath,amssymb}
\usepackage{graphicx}
\usepackage{booktabs}
\usepackage{multirow}
\usepackage{array}
\usepackage{tabularx}
\usepackage{xcolor}
\usepackage{url}
\usepackage{enumitem}
\usepackage{algorithm}
\usepackage{algpseudocode}
\usepackage{colortbl}

\usepackage{makecell}

\usepackage{listings}
\usepackage{xcolor}

\lstdefinestyle{secidjson}{
    basicstyle=\ttfamily\footnotesize\color{black},
    frame=single,
    rulecolor=\color{black},
    framesep=3pt,
    breaklines=true,
    breakatwhitespace=false,
    columns=fullflexible,
    keepspaces=true,
    showstringspaces=false,
    aboveskip=0pt,
    belowskip=0pt
}

\usepackage[
    hidelinks,
    pdfauthor={Md Ajoad Hasan, Dipayan Saha, Upoma Das,
               Sujan Kumar Saha, Mark Tehranipoor,
               Farimah Farahmandi},
    pdftitle={SecID: An LLM-Guided Framework for Threat-Aware Secondary Asset Identification with Quantitative RTL Analysis},
    pdfsubject={SecID preprint}
]{hyperref}
\usepackage[
    a4paper,
    top=19mm,
    bottom=43mm,
    left=15mm,
    right=13mm,
    columnsep=4.10mm
]{geometry}

\usepackage{tikz}
\newcommand{\circled}[1]{\tikz[baseline=(char.base)]{
            \node[shape=circle,draw=black,fill=black, text=white,inner sep=0.5pt] (char) {#1};}}

\newcommand{\sig}[1]{\texttt{\detokenize{#1}}}

\newcolumntype{Y}{>{\raggedright\arraybackslash}X}
\newcolumntype{C}[1]{>{\centering\arraybackslash}p{#1}}

\begin{document}

\newcommand{\papernote}{This paper will be presented at Design, Automation and Test in Europe Conference (DATE) 2027}


\title{%
  \vspace*{-2.0em}%
  {\small\bfseries \papernote\par}%
  \vspace{0.15em}%
  SecID: An LLM-Guided Framework for Threat-Aware
  \underline{Sec}ondary Asset \underline{Id}entification
  with Quantitative RTL Analysis
  \vspace{-0.5em}
}

\author{
\IEEEauthorblockN{\small
Md Ajoad Hasan$^{1}$, Dipayan Saha$^{1}$, Upoma Das$^{1}$,
Sujan Kumar Saha$^{2}$, Mark Tehranipoor$^{1}$ and Farimah Farahmandi$^{1}$}

\IEEEauthorblockA{\small
$^{1}$\textit{Department of Electrical and Computer Engineering, University of Florida, Gainesville, FL, USA}\\
$^{2}$\textit{Department of Electrical and Computer Engineering, Auburn University, Auburn, AL, USA}\\
\{md.hasan, dsaha, upoma.das\}@ufl.edu,
sks0152@auburn.edu,
\{tehranipoor, farimah\}@ece.ufl.edu}

\vspace{-7.5mm}
}

\maketitle

\begin{abstract}
The growing complexity of modern hardware designs is making security assurance increasingly difficult. One of the foundational steps in pre-silicon security verification is identifying security assets. While primary assets are often evident from design functionality, the internal signals that can facilitate their compromise are much less apparent. Complex data- and control-flow dependencies create many such internal security-critical locations, whose exploitation depends on the threat-specific attack mechanism---these are the secondary assets. To precisely identify these, we propose SecID, a multi-agent framework that combines deterministic RTL analysis with bounded dynamic evaluation and LLM reasoning to identify threat-specific secondary assets. Each asset decision is supported by functional influence, threat features, and explicit attack mechanisms. We demonstrate SecID on seven open-source SoC and IP designs, including openMSP430, NEORV32, and CV32E40P. The experimental results are corroborated through comparison with prior work, fault simulation, and formal analysis, and further evaluated through pipeline- and final-stage ablation, threshold-sensitivity studies, and final-stage repeatability analysis. Together, these evaluations show that SecID can identify highly threat-relevant secondary assets with respect to a given primary asset, providing a scalable path toward hardware-security verification.
\end{abstract}

\vspace{-0.65em}

\section{Introduction}
\label{sec:introduction}
\begin{figure}[t]
\centering
\includegraphics[width=0.95\columnwidth]{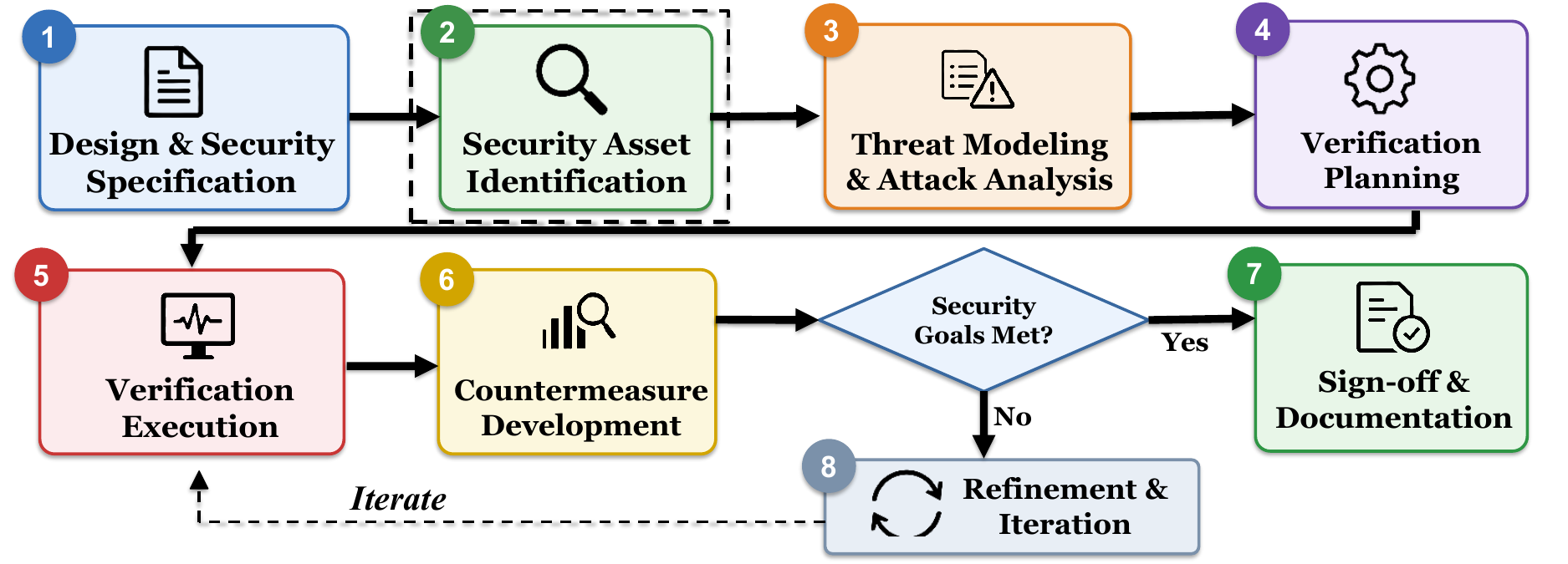}
\vspace{-3.0mm}
\caption{Security-verification Workflow in Hardware Designs.}
\label{fig:security-flow}
\vspace{-6mm}
\end{figure}

\vspace{-0.35em}
Modern systems-on-chip (SoCs) integrate processors, memories, buses, and third-party intellectual-property (IP) blocks via increasingly complex control logic. This high degree of integration introduces security dependencies that may be difficult to recognize at an IP boundary\cite{chen2017soc,knechtel2020eda}. Security verification therefore begins with a fundamental question: \emph{what in the design must be protected?} Asset identification addresses this particular question and guides downstream security tasks, including threat modeling ~\cite{11022932}, property generation ~\cite{tarek2026assertainautomatedsecurityassertion}, and countermeasure design~\cite{saedi,hu2016towards,deutschbein2022toward,witharana2023automated, socurellm}, as summarized in Fig.~\ref{fig:security-flow}.

Of these assets, primary assets are the direct security targets, such as a cryptographic key, program counter (PC), privilege decision, or service-completion signal. Knowing the primary asset tells the verifier what must be protected, but not the locations that can be exploited to compromise it. Its security often depends on surrounding data paths, control signals, state elements, and interfaces that carry, modify, authorize, or control its behavior. An attacker may therefore compromise the primary asset indirectly through these internal elements without attacking the primary asset itself. For example, corrupting a PC-select control can redirect execution without directly modifying the program counter. Such elements are secondary assets~\cite{lasset,saif}. Prior fault-injection attacks demonstrate their security relevance by corrupting internal control, intermediate computations, and verification paths to redirect execution, recover cryptographic keys, or expose protected data~\cite{timmers2016controlling,tang2017clkscrew,murdock2020plundervolt,buhren2021glitch}. Hence, if overlooked, security verification focused only on the primary asset may leave relevant attack paths outside the scope of tests, properties, and countermeasures~\cite{portillo2016building}. Identifying secondary assets therefore extends verification to the internal elements through which the primary asset can be compromised.

However, not every signal structurally connected to a primary asset is necessarily security-relevant. Its significance depends on the threat and the underlying attack mechanism through which it can affect the primary asset. The same signal, for example, may be critical under fault injection but irrelevant to unauthorized access. Structural connectivity alone is therefore insufficient; secondary-asset identification must determine which connected elements can meaningfully contribute to compromise under the considered threat.

Despite their importance, identifying security assets, including secondary assets, has historically relied on manual expert inspection. Only a few recent works have explored automated asset identification~\cite{saha2025sv, 11563216}, and even fewer have focused on secondary assets. For instance, Nath \emph{et al.}~\cite{nath2025} identify potential primary assets using signal-naming and behavioral patterns, while Liu \emph{et al.}~\cite{liu2025structure} classify security assets from RTL data- and control-flow using learned structural signatures. However, both approaches may have limited generalizability when security assets do not exhibit the corresponding naming, behavioral, or structural patterns, and neither provides a threat-specific rationale for the identified assets. LAsset~\cite{lasset} uses LLM-assisted semantic reasoning to provide attack justifications; however, extensive reliance on LLM reasoning can raise reproducibility concerns~\cite{sahaaccess}. SAIF~\cite{saif} is the closest prior work on secondary assets, combining RTL structure with vulnerability analysis. Even there, the threat dependence of secondary assets remains largely underexplored: also, the same primary asset need not have the same secondary-asset set under every threat.

Motivated by the challenges in today’s asset identification approaches, we outline the following research questions (RQs):

\textbf{RQ1:} How can we determine which design elements are secondary assets for a given primary asset?

\textbf{RQ2:} Can we develop an end-to-end methodology to reliably identify secondary assets across diverse design architectures?

\textbf{RQ3:} How can we identify secondary assets while ensuring reproducibility and explainability?

After investigating the answers to the above questions, we propose SecID, a novel LLM-guided agentic framework for systematic secondary asset identification in SoC and IP designs. SecID grounds LLM reasoning in deterministic RTL analysis and bounded threat feature evaluation using EDA tools, providing reproducible evidence for each candidate. Candidate-specific attack mechanisms further complement this evidence with explainable context for the final asset decisions. In summary, the key contributions of this work are as follows:

\begin{itemize}
\item \textbf{Novelty:} To our knowledge, SecID is the first framework that systematically incorporates threat perspectives into secondary asset identification.
\item \textbf{Trustworthiness:} SecID grounds LLM inference in EDA-derived quantitative threat-feature evidence and retains intermediate analysis artifacts for each asset decision, thereby strengthening trustworthiness.


\item \textbf{Explainability:} SecID maps relevant attack mechanisms to the identified secondary assets, providing justifications for the final asset decisions.

\item \textbf{Comprehensiveness:} SecID is evaluated across diverse SoC and IP designs in VHDL, Verilog, and SystemVerilog.

\end{itemize}

The remainder of this paper is organized as follows. Section~\ref{sec:background} presents the background study and rationale. Section~\ref{sec:framework} describes the proposed SecID framework. The experimental results and validation are presented in Section~\ref{sec:experiment}, before concluding the paper in Section~\ref{sec:conclusion}.

\section{Background}
\label{sec:background}
\vspace{-0.25em}
\subsection{Primary vs Secondary Asset}
\vspace{-0.25em}
By definition, a security asset is any hardware component or data element whose protection is essential to preserve the system’s confidentiality, integrity, or availability (CIA). Based on their dependency role in potential security breaches, assets can be categorized as primary or secondary assets~\cite{saif, p3164}.

\begin{figure}[t]
\centering
\includegraphics[width=0.95\columnwidth]{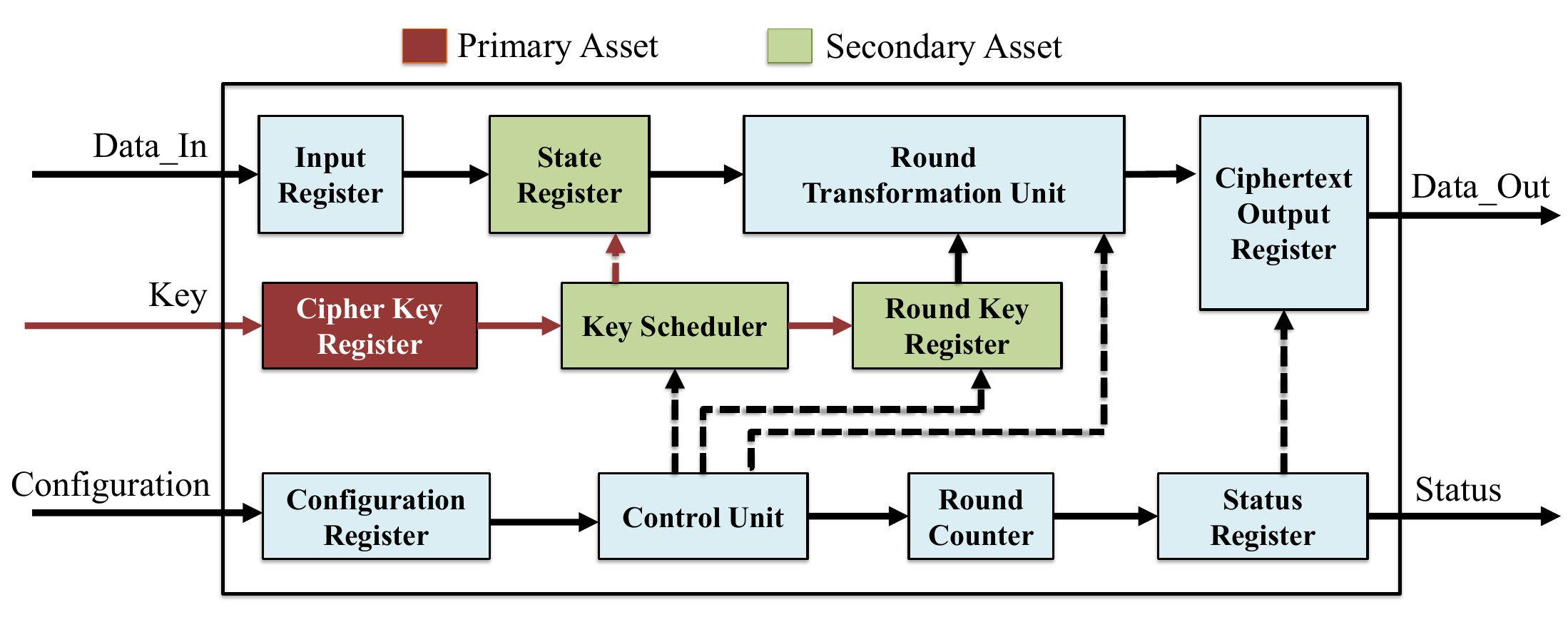}
\vspace{-4.5mm}
\caption{Overview of Asset Types in an AES Encryption Design}
\label{asset_class}
\vspace{-5.0mm}
\end{figure}

\begin{description}[leftmargin=*, labelindent=0.5mm, labelsep=1.5mm]
    \item[$\bullet$] \textbf{Primary Assets}:
    Primary assets are components that serve as the direct targets of an attack. For example, as shown in Fig.~\ref{asset_class}, the encryption key in an AES engine is a primary asset because it is the ultimate target of compromise.

    \item[$\bullet$] \textbf{Secondary Assets}:
    Secondary Assets are components that interact with or facilitate the exposure, corruption, or disruption of primary assets. While not directly critical, their compromise can weaken the overall security of the design by violating the confidentiality, integrity, or availability of the associated primary asset. Examples include system buses, peripheral ports, and internal signals/registers that carry the data of the primary asset, either fully or partially. In AES, elements such as the Encryption/Decryption Engine and the Output Buffer qualify as secondary assets because, though they are not the primary target, their compromise can indirectly expose the encryption key.

\end{description}
\vspace{-0.5em}

\subsection{Threat-Specific Features}
\vspace{-0.25em}

In hardware design, there are four RTL-level threat models that cover the complete CIA objectives: \textbf{(1)} Information Leakage (IL) for confidentiality (C), \textbf{(2)} Unauthorized Access (UA) for both confidentiality (C) and integrity (I), \textbf{(3)} Fault Injection (FI) for integrity (I), and \textbf{(4)} Denial-of-Service (DoS) for availability (A). Our proposed SecID framework considers all four threat models. Each threat employs two complementary RTL-level features that assess how strongly a candidate signal exhibits threat-relevant behavior under the same test scenarios for each candidate; these features are intended for comparative analysis rather than as probabilities of attack success or evidence of exhaustive threat coverage. The two features for each threat are selected to represent distinct aspects of the corresponding attack mechanism while keeping the feature set interpretable and computationally tractable across designs.

\textbf{Information leakage (IL).} IL occurs when confidential information is disclosed through an attacker-observable point. Shannon's mutual information provides an established basis for quantifying statistical dependence~\cite{shannon1948,kastner_jifa}. Therefore, to characterize information leakage, SecID considers \emph{temporal incremental mutual information} and \emph{UOP observability}. \emph{Temporal incremental mutual information} measures how much information about the primary asset accumulates in a candidate signal across cycles, with earlier acquisition weighted more strongly; \emph{UOP observability} asks whether that information reaches a user-specified untrusted observable point (UOP)~\cite{hu2021ift,ardeshiricham2017rtlIFT,chen2016debug}.

\begin{figure*}[t]
  \centering
  \includegraphics[width=1\textwidth]{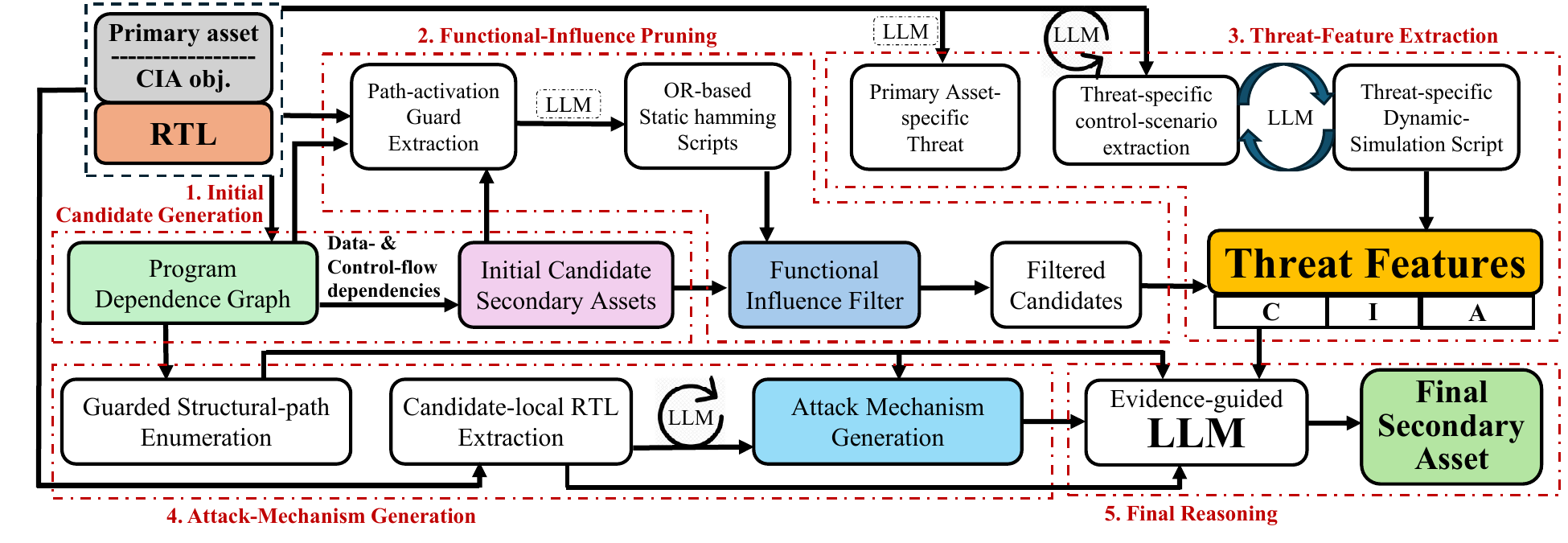}
  \vspace{-8.5mm}
  \caption{Overview of the SecID workflow. Each dashed region represents an agent}
  \label{fig:framework}
  \vspace{-5.5mm}
\end{figure*}

\textbf{Unauthorized access (UA).} UA refers to illicit access to protected resources or operations by bypassing intended authorization through attacker-controllable points. Of the two features, \emph{Bypass sensitivity} measures the fraction of candidate perturbations that turn denial into unauthorized access; \emph{UCP controllability} measures whether user-provided untrusted controllable points (UCPs) can affect the candidate under legal paired scenarios.

\textbf{Fault injection (FI).}
FI perturbs internal data or control state to induce erroneous behavior with adverse system-level impact. For the FI features, fault-injection feasibility (FIF) measures the scenario-balanced fraction of valid fault trials in which the injected fault is activated or consumed by downstream logic, observed through its effect at a downstream point on the candidate-to-primary path or at the primary asset. Fault masking (FM) measures the fraction of activated faults whose effects disappear before corrupting the primary asset.

\textbf{Denial of service (DoS).} DoS maliciously disrupts hardware availability or progress by delaying, blocking, or preventing correct execution. To assess DoS susceptibility, SecID uses \emph{Progress disruption} (PD) to measure the added service delay, finite stall, or failure to recover before or within a bounded post-release window, while \emph{service-rate degradation} (SRD) measures sustained throughput loss under repeated finite-duration blocking of a candidate.

The security impacts characterized by these features are well established in prior hardware-security studies. For example, mutual information has been used in joint information-flow tracking~\cite{kastner_jifa,hu2017criticalpath}. Similarly, fault injection studies commonly consider fault activation, propagation, and masking~\cite{chen2021voltpillager,barenghi2012fault}. DoS studies discuss effects such as service delay ~\cite{rajan2020sectar,prasad2017runtime,fang2013robustness}. Hence, SecID does not treat the underlying phenomena as novel; rather, it adapts them into candidate-level RTL features for threat-specific secondary-asset identification, which, to our knowledge, has not previously been explored.

\vspace{-0.20em}
\section{SecID Framework}
\label{sec:framework}
\vspace{-0.15em}
Identifying secondary assets from RTL, as posed in \textbf{RQ1--RQ3} in Section~I, requires a solution with several key stages. It must first extract the broad set of signals structurally connected to the primary asset through RTL analysis and then selectively filter this set based on functional influence. The retained candidates must next be assessed using threat-specific quantitative evidence, while the surrounding design context must also be interpreted to determine plausible ways in which each candidate can contribute to compromise of the primary asset. The final decision should therefore be guided by both threat evidence and semantic attack context, thereby enhancing reproducibility and explainability. SecID realizes these requirements through a multi-agent approach that combines EDA-based RTL analysis with LLM reasoning. The EDA-based stages generate structural, functional, and quantitative threat evidence, while LLMs are used where design-dependent interpretation is required, including threat-scenario construction, EDA-script generation and refinement, attack-mechanism generation, and final asset decisions. To support scalability to larger designs, only the RTL segment local to each primary--candidate pair is extracted preserving its hierarchy.

As shown in Fig.~\ref{fig:framework}, SecID operates through five interacting agents: \circled{\small 1} Initial Candidate Generation, \circled{\small 2} Functional-Influence Pruning, \circled{\small 3} Threat-Feature Extraction, \circled{\small 4} Attack-Mechanism Generation, and \circled{\small 5} Final LLM Reasoning. This modular agentic workflow is designed to enable scalable and traceable secondary asset identification across diverse SoC and IP designs.

\vspace{-2mm}
\subsection{Agent 1: Initial Candidate Generation}
\vspace{-1mm}
Agent~1 elaborates the RTL into an inter-module Program Dependence Graph (PDG) containing both data and control dependencies~\cite{ferrante1987pdg}. For confidentiality, this agent follows primary-asset fan-out; for integrity and availability, it follows fan-in. Cross-hierarchy representations of the same signal and simple assignment aliases are de-duplicated. For openMSP430 \sig{frontend_0.pc}, the resulting candidate set includes next-PC data signals together with state, interrupt, fetch, and debug-write controls. Agent~1's steps are summarized in Algorithm~\ref{alg:secid} (lines~1--3).
\vspace{-3.0mm}

\subsection{Agent 2: Functional-Influence Pruning}
\vspace{-1.5mm}
Agent~2, as presented in Algorithm~\ref{alg:secid} (lines~4--11), serves as a conservative screening step that filters out candidates with consistently weak functional influence while retaining those with stronger or unresolved influence. This allows later agents to operate on a reduced set of functionally relevant candidates rather than the full candidate universe.

\subsubsection{\textbf{Path-activation guard extraction}}
Branch conditions, mux selects, FSM states, and related control predicates are extracted deterministically from the PDG to identify when primary--candidate paths become active. These predicates are stored in JSON for subsequent functional-influence analysis.

\subsubsection{\textbf{Boolean influence analysis}}
Each local RTL cone is reduced to Boolean-level operators, and conservative transfer rules are assigned to logic, arithmetic, comparisons, and multiplexers. For example, a mux data input contributes only when its branch guard is active, while arithmetic and logical operators propagate per-bit influence according to their Boolean behavior. Thus, strongly connected components are condensed, and influence is propagated without running an RTL testbench, under the path-activation guards extracted above. Reconvergent contributions are combined using an OR-style composition so that multiple alternative paths contribute without canceling one another.

\subsubsection{\textbf{Static Hamming approximation}}
Using the Boolean-influence values in the previous stage, four normalized metrics summarize how strongly each candidate can functionally relate to the primary asset. $H_{\mathrm{agg}}$ captures the aggregate expected Hamming influence across multiple reconvergent paths; $H_{\mathrm{path}}$ the strongest ordinary bit-to-bit path influence; $H_{\mathrm{guard}}$ the strongest influence under an activating guard; and $H_{\mathrm{legal}}$ the influence under structurally valid mutually exclusive activation conditions, with \(H_{\mathrm{legal}}\) used only for fan-in analysis. These complementary views account for different path behaviors that may be missed by a single metric. A candidate is retained if any of the views reaches a fixed design-independent threshold.

\begin{table}[t]
\caption{Representative Static Functional-Influence Values}
\vspace{-2mm}
\label{tab:static}
\centering
\fontsize{9.5}{9.5}\selectfont
\renewcommand{\arraystretch}{0.95}
\setlength{\tabcolsep}{2.0pt}
\setlength{\aboverulesep}{0.2ex}
\setlength{\belowrulesep}{0.2ex}
\setlength{\cmidrulesep}{0.15ex}

\resizebox{\columnwidth}{!}{%
\begin{tabular}{
@{}
>{\raggedright\arraybackslash}p{1.95cm}
>{\raggedright\arraybackslash}p{2.10cm}
>{\raggedright\arraybackslash}p{3.35cm}
>{\centering\arraybackslash}p{1.05cm}
>{\centering\arraybackslash}p{1.05cm}
>{\centering\arraybackslash}p{1.05cm}
>{\centering\arraybackslash}p{1.05cm}
@{}
}

\toprule
\noalign{\vskip 2pt}

Design &
Primary Asset &
Candidate &
$H_{\mathrm{agg}}$ &
$H_{\mathrm{path}}$ &
$H_{\mathrm{guard}}$ &
$H_{\mathrm{legal}}$ \\

\midrule

\multirow[c]{2}{=}{OpenMSP430} &
\multirow[c]{2}{=}{\path{omsp_frontend.pc}} &
\path{frontend_0.pc_sw}
& 0.0313
& 0.500
& 1.000
& 0
\\

\arrayrulecolor{black!65}
\cmidrule(lr){3-7}
\arrayrulecolor{black}

& &
\path{inst_src}
& 0.0178
& 0.00391
& 0.00781
& 0.250
\\

\midrule

\multirow[c]{2}{=}{AES Core} &
\multirow[c]{2}{=}{\path{aes_cipher_top.key}} &
\makecell[l]{
    \path{aes_cipher_top.u0.}\\[-1pt]
    \path{subword}
}
& 1.000
& 0.500
& 1.000
& --
\\

\arrayrulecolor{black!65}
\cmidrule(lr){3-7}
\arrayrulecolor{black}

& &
\path{aes_cipher_top.sa00}
& 0.625
& 0.250
& 0.500
& --
\\

\midrule

\multirow[c]{2}{=}{NEORV32} &
\multirow[c]{2}{=}{\path{ctrl_o.lsu_priv}} &
\makecell[l]{
    \path{neorv32_cpu_control.}\\[-1pt]
    \path{ctrl.csr_we}
}
& 0.4375
& 0.250
& 0.500
& 0
\\

\arrayrulecolor{black!65}
\cmidrule(lr){3-7}
\arrayrulecolor{black}

& &
\makecell[l]{
    \path{neorv32_cpu_control.}\\[-1pt]
    \path{csr.dcsr_cause}
}
& $4.65\!\times\!10^{-9}$
& $7.45\!\times\!10^{-9}$
& $1.49\!\times\!10^{-8}$
& 0
\\

\bottomrule
\end{tabular}%
}
\vspace{-1.85em}
\end{table}

\vspace{-2.5mm}

\subsection{Agent 3: Threat-Feature Extraction}
For each candidate retained by Agent~2, Agent~3 calculates a set of threat features via bounded dynamic RTL simulation, as shown in Algorithm~\ref{alg:secid} (lines~12--15).

\subsubsection{\textbf{Primary asset-specific threat modeling}}
An LLM sub-agent first identifies the threat(s) relevant to the primary asset based on its CIA objective and RTL context. Only the features corresponding to the threat(s) are considered subsequently.

\subsubsection{\textbf{Threat-feature score calculation}}

Given the selected threat and RTL context, the LLM generates targeted scenario classes for each feature and simulation scripts through in-context learning. A threat scenario specifies a concrete RTL operating condition under which the selected threat can be exercised and the corresponding candidate behavior can be measured. For example, for information leakage of \sig{r4}, a valid debug read of \sig{r4} while monitoring \sig{dbg_uart_txd} defines a scenario for testing whether candidate-dependent information can propagate to the UOP.
Multiple independent LLM runs are first unioned and semantically de-duplicated; the resulting scenario pool is then iteratively fed back to the LLM to identify any remaining scenarios until the fraction of newly added classes falls below a fixed threshold, thereby improving scenario coverage. Based on these scenarios, the generated testbenches are iteratively refined using compilation and simulation feedback through the LLM. However, candidates disabled by RTL parameterization are marked as configuration-inactive.

\begin{algorithm}
\caption{Threat-aware Secondary Asset Identification}
\label{alg:secid}
\begin{algorithmic}[1]
\scriptsize

\Require RTL $D$, Primary asset $p$, CIA obj. $o$,
UOP/UCP $U$, Threshold $\tau$
\Ensure Secondary assets $S$ with Threat feat. evidence and Attack mechanism

\Statex \textbf{/* Agent 1: PDG-based candidate generation */}
\State $G \gets \mathrm{PDGElaboration}(D)$
\State $C_0 \gets \mathrm{DeAlias}
       (\mathrm{DirectedReach}(G,p,o))$
\State $D_{\mathrm{loc}} \gets
       \mathrm{LocalRTL}(D,G,p,C_0)$

\Statex \textbf{/* Agent 2: Functional-influence pruning */}
\State $\Gamma \gets
       \mathrm{GuardExtraction}(D_{\mathrm{loc}},G)$
\ForAll{$c \in C_0$}
    \State $b_c \gets \mathrm{BooleanInfluence}(p,c,\Gamma)$
    \State $h_c \gets \mathrm{StaticHamming}(b_c,\Gamma)$
    \If{$\max(h_c) \ge \tau$ or $h_c$ is unresolved}
        \State add $c$ to $C$
    \EndIf
\EndFor

\Statex \textbf{/* Agent 3: Threat-feature extraction */}
\State $T \gets
       \mathrm{ThreatSelection}(p,o,D_{\mathrm{loc}})$
\State $Q \gets
       \mathrm{ThreatScenarioGeneration}(T,U,D_{\mathrm{loc}})$
\State $B \gets
       \mathrm{EDAScriptRefinement}(Q,D_{\mathrm{loc}})$
\State $F \gets
       \mathrm{DynamicFeatureSimulation}(p,C,T,U,B)$

\Statex \textbf{/* Agent 4: Attack-mechanism generation */}
\State $K \gets \mathrm{DependencyClusters}(C,G)$
\State $(R,P) \gets
       \mathrm{LocalRTLAndPaths}(D_{\mathrm{loc}},G,p,K)$
\State $A \gets
       \mathrm{LLMAttackMechanisms}(K,T,R,P)$

\Statex \textbf{/* Agent 5: Final LLM reasoning */}
\State $S \gets
       \mathrm{LLMJointReasoning}(C,F,A,R,P)$

\State \Return $S$
\end{algorithmic}
\makeatletter
\appto{\@fs@post}{\vspace{-8pt}}
\makeatother
\end{algorithm}

Constrained bounded dynamic simulation is used to assess relative threat relevance without implying exhaustive attack coverage. Candidates within the same experiment are evaluated under common scenarios and observation bounds. A missing witness or insufficiently activated scenario is marked unresolved rather than assigned a low score, and is passed to Agent~5. Below, $c$ denotes a candidate, $s\in\mathcal{S}_c$ a legal scenario, and $\mathbf{1}[\cdot]$ the indicator function.

For \textbf{IL}, let $\Delta I_{c,s}(t)$ denote the incremental mutual information about primary asset $P$ observed at candidate $c$ in cycle $t$, with $T_s$ the observation length. Only scenarios with $H(P|s)>0$ are included. For UOP set $\mathcal{O}$, $\mathrm{Obs}(c,o,s)$ denotes a valid candidate-to-UOP observability witness.

\vspace{-0.65em}
{\footnotesize
\begin{equation}
\begin{aligned}
F_{\mathrm{TIMI}}(c)
 &= \max_{\substack{s\in\mathcal{S}_c\\H(P|s)>0}}
    \sum_{t=1}^{T_s}
    \frac{T_s-t+1}{T_s}
    \frac{\Delta I_{c,s}(t)}{H(P|s)},\\
F_{\mathrm{UOP}}(c)
 &= \frac{1}{|\mathcal O|}
    \sum_{o\in\mathcal O}
    \mathbf{1}[\exists s:\mathrm{Obs}(c,o,s)] .
\end{aligned}
\vspace{-0.5em}
\end{equation}
}

For \textbf{UA}, bypass sensitivity is computed over candidate bits $b$ having at least one valid baseline-denied perturbation and maximized across such bits. For UCP set $\mathcal{U}$, $u\leadsto c$ denotes a valid paired controllability witness.

\vspace{-1.0em}
{\footnotesize
\begin{equation}
\begin{aligned}
BS(c)
 &= \max_{\substack{b\\N_{\mathrm{valid\,deny}}(c,b)>0}}
    \frac{N_{\mathrm{deny}\rightarrow\mathrm{UA}}(c,b)}
         {N_{\mathrm{valid\,deny}}(c,b)},\\
F_{\mathrm{UCP}}(c)
 &= \frac{1}{|\mathcal U|}
    \sum_{u\in\mathcal U}
    \mathbf{1}[\exists s:u\leadsto c].
\end{aligned}
\vspace{-0.5em}
\end{equation}
}

For \textbf{FI}, a fault is considered propagated when its effect reaches an independently monitored downstream point on the candidate-to-primary path or directly corrupts the primary asset. FIF is the scenario-balanced fraction of valid fault trials showing such propagation, while FM measures the fraction of activated faults masked before primary-asset corruption.

\vspace{-0.55em}
{\footnotesize
\begin{equation}
\begin{aligned}
FIF(c)
 &= \frac{1}{|\mathcal S_c|}
    \sum_s
    \frac{N_{\mathrm{act}}(c,s)}
         {N_{\mathrm{valid\ FI}}(c,s)},\\
FM(c)
 &= 1-\frac{1}{|\mathcal S_c|}
    \sum_s
    \frac{N_{\mathrm{primary}}(c,s)}
         {N_{\mathrm{act}}(c,s)} .
\end{aligned}
\vspace{-0.5em}
\end{equation}
}

Here, $N_{\mathrm{act}}$ counts valid trials in which the injected fault is activated or consumed, identified by a fault-induced effect at a monitored downstream point or at the primary asset, while $N_{\mathrm{primary}}$ counts trials that corrupt the primary asset.

For \textbf{DoS}, $W_A$ and $W_R$ denote the attack and post-release recovery windows, respectively; $\Delta T_j$ is the added service delay, and $R_j$ is the release-to-recovery latency for trial $j$.

\vspace{-1.15em}
{\footnotesize
\begin{equation}
d_{\mathrm{PD}}(j)=
\begin{cases}
\dfrac{\Delta T_j}{W_A+W_R},
& \text{completion during } W_A,\\[4pt]
\dfrac{W_A+R_j}{W_A+W_R},
& \text{no completion in }W_A,\ \text{recovery in }W_R,\\[4pt]
1,
& \text{no recovery within }W_R .
\end{cases}
\vspace{-0.5em}
\end{equation}
}

Here, $\Omega_{c,s}$ denotes the valid blocking trials, while $N_{\mathrm{base}}$ and $N_{\mathrm{blk}}$ denote legitimate service completions without and with repeated blocking, respectively.

\vspace{-0.45em}
{\footnotesize
\begin{equation}
\begin{aligned}
PD(c)
 &= \frac{1}{|\mathcal S_c|}
    \sum_s \frac{1}{|\Omega_{c,s}|}
    \sum_{j\in\Omega_{c,s}} d_{\mathrm{PD}}(j),\\
SRD(c)
 &= \frac{1}{|\mathcal S_c|}
    \sum_s
    \max\!\left(
    0,\,
    1-\frac{N_{\mathrm{blk}}(c,s)}
            {N_{\mathrm{base}}(s)}
    \right).
\end{aligned}
\vspace{-0.5em}
\end{equation}
}

\begin{table*}[t]
\caption{Representative Threat-Feature Values}
\vspace{-2mm}
\label{tab:features}
\centering
\scriptsize
\renewcommand{\arraystretch}{0.90}
\setlength{\tabcolsep}{1.05pt}
\setlength{\aboverulesep}{0.20ex}
\setlength{\belowrulesep}{0.20ex}
\setlength{\cmidrulesep}{0.15ex}

\begin{tabularx}{\textwidth}{
@{}
>{\centering\arraybackslash}m{1.50cm}
>{\centering\arraybackslash}m{1.40cm}
>{\centering\arraybackslash}m{2.85cm}
>{\centering\arraybackslash}m{1.30cm}
>{\centering\arraybackslash}m{1.30cm}
>{\centering\arraybackslash}m{1.10cm}
>{\centering\arraybackslash}m{1.30cm}
>{\centering\arraybackslash}m{1.30cm}
>{\centering\arraybackslash}m{1.30cm}
>{\centering\arraybackslash}m{1.30cm}
>{\centering\arraybackslash}m{1.30cm}
>{\centering\arraybackslash}m{1.55cm}
@{}}

\toprule

& & &
\multicolumn{2}{c}{\raisebox{-1.5pt}{Information Leakage}} &
\multicolumn{2}{c}{\raisebox{-1.5pt}{Unauthorized Access}} &
\multicolumn{2}{c}{\raisebox{-1.5pt}{Fault Injection}} &
\multicolumn{2}{c}{\raisebox{-1.5pt}{Denial-of-Service}} &
\\[2pt]

\cmidrule(lr){4-5}
\cmidrule(lr){6-7}
\cmidrule(lr){8-9}
\cmidrule(lr){10-11}

\noalign{\vskip -3pt}

\makecell[c]{Design} &
\makecell[c]{Primary\\Asset} &
\makecell[c]{Candidate} &
\makecell[c]{Incremental\\Mutual\\Information} &
\makecell[c]{UOP\\Observ-\\ability} &
\makecell[c]{UCP\\Control-\\lability} &
\makecell[c]{Bypass\\Sensitivity} &
\makecell[c]{Fault Injection\\Feasibility} &
\makecell[c]{Fault\\Masking} &
\makecell[c]{Progress\\Disruption} &
\makecell[c]{Service Rate\\Degradation} &
\makecell[c]{Context} \\
\midrule

\multirow[c]{2}{=}{OpenMSP430} &
\multirow[c]{2}{=}{\sig{r4}} &
\path{execution_unit_0.alu_out} &
0.958 &
1 &
$\times$ &
$\times$ &
$\times$ &
$\times$ &
$\times$ &
$\times$ &
\multirow[c]{2}{1.45cm}{
    \raisebox{1.5pt}{
    \makecell[c]{UOP:\\[-2pt]
    \sig{dbg_uart_}\\[-2pt]
    \sig{txd}}}
}
\\

\arrayrulecolor{black!65}
\cmidrule(lr){3-11}
\arrayrulecolor{black}

& &
\path{frontend_0.irq_all} &
0.0495 &
1 &
$\times$ &
$\times$ &
$\times$ &
$\times$ &
$\times$ &
$\times$ &
\\

\midrule

\multirow[c]{2}{=}{CV32E40P} &
\multirow[c]{2}{=}{\sig{csr_we_int}} &
\makecell[c]{
    \path{core_i.id_stage_i.}\\[-1pt]
    \path{deassert_we}
} &
$\times$ &
$\times$ &
0.927 &
0.161 &
0.532 &
0.059 &
$\times$ &
$\times$ &

\multirow[c]{2}{1.45cm}{
    \raisebox{1.5pt}{
        \makecell[c]{
            UCPs:\\[-2pt]
            \sig{instr_rdata}\\[-2pt]
            \sig{_i}, \sig{debug}\\[-2pt]
            \sig{_req_i}
        }
    }
}
\\

\arrayrulecolor{black!65}
\cmidrule(lr){3-11}
\arrayrulecolor{black}

& &
\makecell[c]{
    \path{core_i.cs_registers_i.}\\[-1pt]
    \path{mhpmcounter_write_upper}
} &
$\times$ &
$\times$ &
0 &
0 &
0.085 &
1.000 &
$\times$ &
$\times$ &
\\

\midrule

\multirow[c]{2}{=}{AXI adapter} &
\multirow[c]{2}{=}{\sig{valid_o}} &
\path{state_d} &
$\times$ &
$\times$ &
$\times$ &
$\times$ &
$\times$ &
$\times$ &
0.762 &
0.783 &
\\

\arrayrulecolor{black!65}
\cmidrule(lr){3-11}
\arrayrulecolor{black}

& &
\path{axi_resp_i.b_valid} &
$\times$ &
$\times$ &
$\times$ &
$\times$ &
$\times$ &
$\times$ &
0.319 &
0.994 &
\\

\bottomrule
\end{tabularx}
\vspace{-1em}
\end{table*}

\vspace{-0.5em}
\subsection{Agent 4: Attack-Mechanism Generation}
\vspace{-0.25em}
Unlike the threat scenarios in Agent~3, the attack mechanisms generated here explain how a specific candidate can participate in compromising the primary asset. For example, when MPRV is active, \sig{mstatus_mpp} determines load/store privilege; corrupting it can therefore alter the effective privilege carried by \sig{lsu_priv}.

Taking the filtered candidates from Agent~2, this agent operates independently of Agent~3 to provide attack context. RTL-proximal candidates are clustered in groups of 3--5, and each API call includes PDG-extracted structural paths and localized RTL blocks for the cluster. The LLM then identifies the required preconditions, propagation route, and security consequence as structured JSON output, as shown in Algorithm~\ref{alg:secid} (lines~16--18). The resulting attack-mechanism pool is iteratively enriched, similar to threat-scenario generation.

\vspace{-0.5em}
\subsection{Agent 5: Final LLM Reasoning}
\vspace{-0.25em}
The final agent receives the localized RTL, guarded paths, threat-feature configurations and values, and candidate-specific attack mechanisms as input. It is instructed to retain a signal only when these artifacts consistently support the same threat-relevant relation, as presented in Algorithm~\ref{alg:secid} (lines~19--20). Quantitative evidence suppresses candidates with consistently weak measured behavior, while the attack context helps resolve cases that a scalar score alone cannot explain. Unresolved evidence is considered conservatively rather than treated as failure. The final output is JSON for each primary-asset--threat pair, as shown in Fig.~\ref{fig:json-output}.

\begin{figure}[t]
\centering
\begin{minipage}{0.98\columnwidth}

\begin{lstlisting}[
    style=secidjson,
    basicstyle=\ttfamily\fontsize{6.5}{6.5}\selectfont,
    escapeinside={(*@}{@*)}
]
{
  (*@\textcolor{green!45!black}{\ttfamily "Name"}@*): "frontend_0.i_state",
  (*@\textcolor{green!45!black}{\ttfamily "Structural Path"}@*):
    "frontend_0.i_state -> fetch/decode sequencing -> PC-update selection -> omsp_frontend.pc",
  (*@\textcolor{green!45!black}{\ttfamily "Attack Scenarios"}@*): [
    "A transient fault on frontend_0.i_state can perturb fetch/decode 
     sequencing -> PC-update selection, producing an unintended value, enable, 
     or transition at omsp_frontend.pc."
  ],
  (*@\textcolor{green!45!black}{\ttfamily "Normalized Evidence"}@*): {
    (*@\textcolor{green!45!black}{\ttfamily "fault\_injection\_feasibility"}@*): 0.958333333333,
    (*@\textcolor{green!45!black}{\ttfamily "fault\_masking"}@*): 0.0416666666667
  },
  (*@\textcolor{green!45!black}{\ttfamily "Justification"}@*):
    "The instruction FSM selects fetch and interrupt PC-update behavior, so a 
     state fault can place the PC on the wrong transition.",
  (*@\textcolor{green!45!black}{\ttfamily "decision"}@*):
    "retain"
}
\end{lstlisting}

\end{minipage}

\vspace{-0.3em}
\caption{Example SecID secondary-asset output for
\sig{frontend_0.pc}.}
\label{fig:json-output}
\vspace{-1.5em}
\end{figure}

\begin{table*}[t]
\caption{Summarized Asset Results of SecID}
\vspace{-2mm}
\label{tab:results}
\centering
\footnotesize
\renewcommand{\arraystretch}{0.93}
\setlength{\tabcolsep}{2.3pt}
\setlength{\aboverulesep}{0.3ex}
\setlength{\belowrulesep}{0.3ex}
\setlength{\cmidrulesep}{0.2ex}

\begin{tabularx}{\textwidth}{
@{}
>{\raggedright\arraybackslash}p{1.55cm}
>{\raggedright\arraybackslash}p{3.00cm}
>{\centering\arraybackslash}p{1.65cm}
>{\centering\arraybackslash}p{0.65cm}
>{\centering\arraybackslash}p{0.70cm}
>{\centering\arraybackslash}p{0.70cm}
>{\centering\arraybackslash}p{1.45cm}
>{\centering\arraybackslash}p{0.65cm}
>{\raggedright\arraybackslash}X
@{}}

\toprule
Design &
Primary Asset &
\makecell{Security\\Objective} &
Threat &
Initial &
Filtered &
\makecell{Threat\\Scenarios} &
Final &
Representative Final Secondary Assets \\
\midrule

\multirow[c]{3}{*}{OpenMSP430}
&
\path{omsp_frontend.pc}
&
Integrity
&
FI
&
453 & 150 & 77 & 19
&
\sig{i_state}, \sig{e_state}, \sig{pc_sw_wr},
\sig{reg_pc_call}
\\

\arrayrulecolor{black!65}
\cmidrule(lr){2-9}
\arrayrulecolor{black}

&
\path{execution_unit_0.register_file_0.r4}
&
Confidentiality
&
IL
&
441 & 93 & 40 & 10
&

\sig{reg_src}, \sig{op_src}, \sig{alu_out},
\sig{reg_incr_val}, \sig{mdb_out_nxt},
\sig{mem_data}, \sig{xfer_buf_nxt}

\\

\arrayrulecolor{black!65}
\cmidrule(lr){2-9}
\arrayrulecolor{black}

&
\path{omsp_frontend.exec_done}
&
Availability
&
DoS
&
453 & 65 & 62 & 11
&
\sig{e_state}, \sig{decode}, \sig{exec_dst_wr},
\sig{inst_dext_en}, \sig{exec_dext_rdy}, \sig{exec_jmp}
\\

\midrule

\multirow[c]{2}{*}{NEORV32}
&
\multirow[c]{2}{2.85cm}{\path{ctrl_o.lsu_priv}}
&
\multirow[c]{2}{*}{Integrity}
&
UA
&
\multirow[c]{2}{*}{175}
&
\multirow[c]{2}{*}{52}
&
52
&
12
&
\sig{prv_level}, \sig{mstatus_mpp},
\sig{ctrl_nxt.csr_we}
\\

\arrayrulecolor{black!65}
\cmidrule(lr){4-4}
\cmidrule(lr){7-9}
\arrayrulecolor{black}

&
&
&
FI
&
&
&
52
&
8
&
\sig{mstatus_mpp}, \sig{dcsr_prv},
\sig{env_exit}, \sig{exec.ir}
\\

\midrule

\multirow[c]{2}{*}{CV32E40P}
&
\multirow[c]{2}{3.00cm}
{\path{core_i.cs_registers_i.csr_we_int}}
&
\multirow[c]{2}{*}{Integrity}
&
UA
&
\multirow[c]{2}{*}{494}
&
\multirow[c]{2}{*}{37}
&
164
&
4
&
\sig{deassert_we}, \sig{decoder_i.csr_op}
\\

\arrayrulecolor{black!65}
\cmidrule(lr){4-4}
\cmidrule(lr){7-9}
\arrayrulecolor{black}

&
&
&
FI
&
&
&
42
&
7
&
\sig{data_misaligned}, \sig{csr_op},
\sig{id_ready}, \sig{ex_ready}
\\

\midrule

\multirow[c]{2}{*}{APB GPIO}
&
\multirow[c]{2}{2.85cm}{\path{gpio_dir}}
&
\multirow[c]{2}{*}{Integrity}
&
UA
&
\multirow[c]{2}{*}{9}
&
\multirow[c]{2}{*}{9}
&
20
&
3
&
\sig{PADDR[6:2]}, \sig{PWDATA[0]},
\sig{r_gpio_lock[0]}
\\

\arrayrulecolor{black!65}
\cmidrule(lr){4-4}
\cmidrule(lr){7-9}
\arrayrulecolor{black}

&
&
&
FI
&
&
&
34
&
6
&
\sig{PENABLE}, \sig{r_gpio_lock},
\sig{PWDATA}, \sig{PSEL}
\\

\midrule

AXI adapter
&
\path{valid_o}
&
Availability
&
DoS
&
16 & 16 & 23 & 9
&
\sig{ar_ready}, \sig{aw_ready},
\sig{state_q}, \sig{cnt_q}
\\

\midrule

SHA-3
&
\path{keccak_in}
&
Confidentiality
&
IL
&
144 & 144 & 40 & 6
&
\sig{v1}, \sig{out}, \sig{round_in}
\\

\midrule

AES core
&
\path{key}
&
Confidentiality
&
IL
&
62 & 62 & 28 & 20
&
\sig{w[0]}, \sig{sa00}, \sig{sa33}
\\

\bottomrule
\end{tabularx}
\vspace{-1.85em}
\end{table*}

\section{Experimental Results and Evaluation}
\label{sec:experiment}
\vspace{-0.15em}
\subsection{Experimental Setup}
\vspace{-0.15em}
We evaluated SecID on 7 open-source RTL designs: 3 processors/cores (openMSP430, NEORV32, and CV32E40P), 2 peripherals (APB GPIO and AXI adapter), and 2 cryptographic IPs (SHA-3 and AES core)~\cite{openmsp430,neorv32,cv32e40p,opencores}. For each design, we selected 1--3 representative primary assets that provide distinct security-critical functions, such as control flow, privilege enforcement, service progress, and cryptographic secrets, so that the case studies cover diverse architectural roles and attack mechanisms.

The methodology employs PyVerilog and NetworkX for PDG extraction, Yosys for Boolean-influence analysis, and Icarus Verilog and Synopsys VCS for bounded dynamic simulation. GPT-5.6-Sol is used through the API for scenario extraction, script generation and refinement, attack-mechanism generation, and final reasoning. Python is used throughout the workflow to connect these stages, perform numerical analysis, and manage intermediate outputs. Across all experiments, Agent~2 uses a fixed functional-influence threshold of $\tau=0.05$.

\vspace{-0.5em}

\subsection{Processor and IP Case Studies}
\begin{figure*}[t]
\centering
\includegraphics[width=0.95\textwidth]{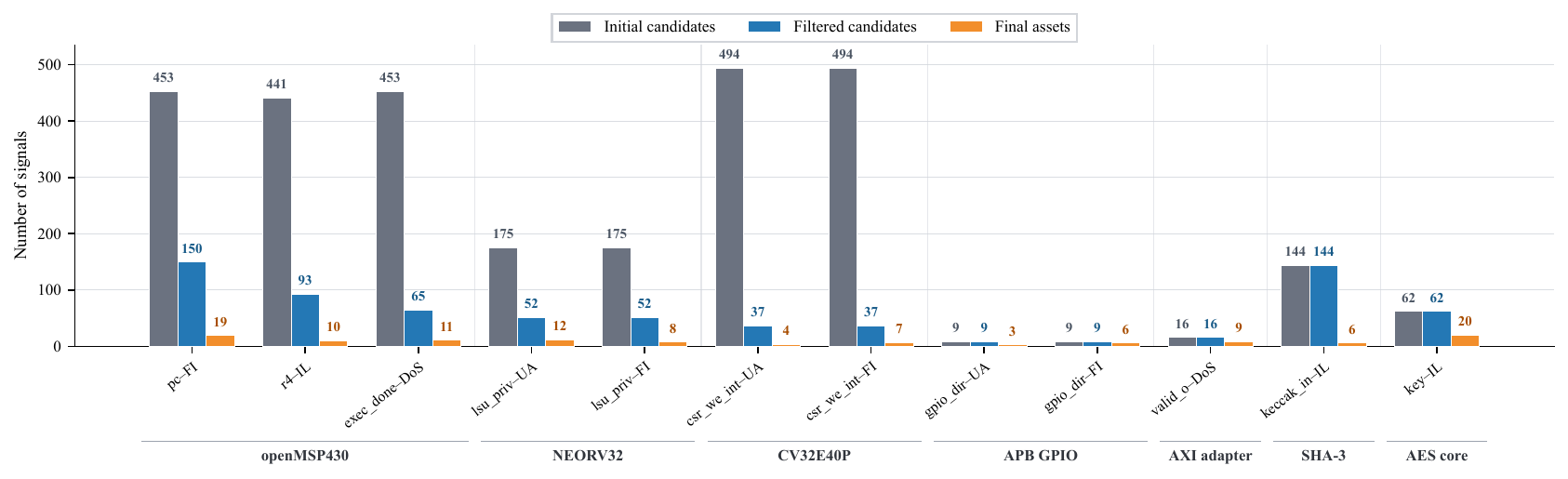}
\vspace{-3.5mm}
\caption{Summarized statistics of candidate reduction across all case studies.}
\label{fig:stats}
\vspace{-1.5em}
\end{figure*}

\vspace{-0.25em}
Tables~\ref{tab:static}--\ref{tab:results} summarize SecID's progression from functional pruning through threat-feature extraction to final asset identification. Fig.~\ref{fig:stats} visualizes the candidate statistics across all 12 cases. Results are available online.\footnote{\url{https://github.com/Ajoad/SecID-Secondary-Assets}}

In \textbf{SHA-3} and \textbf{AES}, the iterative and highly reconvergent datapaths make Agent~2 pruning less selective because primary-asset influence rapidly propagates through the recurrent round/state logic. Consequently, all candidates in both designs pass the functional-influence threshold.

\vspace{-0.5em}
\subsection{Validation}
\vspace{-0.25em}
In the absence of a comprehensive benchmark for secondary-asset identification, SecID is validated through a number of complementary checks: comparison with SAIF and Z01X fault simulation, ablation against independently curated expert reference sets, and JasperGold formal analysis.

\subsubsection{Consistency with SAIF and Fault Simulation}
We first compare the openMSP430 \sig{pc} functional-influence screening against the Pearson Correlation Coefficient (PCC) examples reported in SAIF~\cite{saif}. Agent~2 of SecID retains all seven high-PCC examples and prunes four of the six low-PCC examples, with \sig{rd_wr} excluded because it was not found in the available RTL.
In SAIF, PCC provides a non-directional measure of long-window signal-activity correlation. Therefore, this comparison is used only as an independent consistency check for functional pruning in Agent~2 of SecID.

The functionally filtered candidates of the same openMSP430 \sig{pc} were further evaluated through stuck-at fault simulation using Synopsys VC Z01X~\cite{z01x}. FIF versus the Z01X fault-attempt rate gives a Spearman correlation of $\rho=0.444$, while $1-\mathrm{FM}$ versus the conditional PC-impact rate gives $\rho=0.857$. More importantly, the combined analytical fault susceptibility, $\mathrm{FIF}(1-\mathrm{FM})$, shows strong rank agreement ($\rho=0.888$) with the balanced overall PC-impact rate ($p\approx4.3\times10^{-51}$; bootstrap 95\% CI $[0.824,0.935]$). This agreement serves as an external assessment of Agent~3's FI feature construction.

\subsubsection{Pipeline and Final-Stage Ablation}
To examine how SecID progressively refines the candidate set and how each type of evidence contributes to the final decision, we compare the outputs of successive pipeline stages against reference sets. These sets were independently curated by experts with RTL and hardware-security expertise; a signal was labeled as a secondary asset only when RTL analysis established a feasible structural and functional path through which the specified threat could be realized. Table~\ref{tab:ablation} presents this comparison for four case studies. At the final reasoning stage, RTL-only and attack-only configurations tend to over-select plausible candidates, with RTL-only showing slightly weaker overall performance. Nonetheless, the attack-only results indicate that attack-mechanism context alone may be insufficient to distinguish all threat-relevant candidates. Feature-only reasoning is more selective but still underperforms the full SecID flow, supporting the complementary roles of quantitative threat evidence and attack-mechanism context.

\begin{table}[t]
\caption{Pipeline and final-stage ablation results across four cases (TP/FP/FN). 
{\scriptsize TP: true positive; FP: false positive; FN: false negative.}}
\vspace{-2mm}
\label{tab:ablation}
\centering
\scriptsize
\renewcommand{\arraystretch}{1.08}
\setlength{\tabcolsep}{1.4pt}
\resizebox{\columnwidth}{!}{%
\begin{tabular}{@{}lcccccc@{}}
\toprule
Configuration &
AES &
OpenMSP430 &
APB GPIO &
OpenMSP430 &
\multicolumn{2}{c}{Aggregate} \\
\cmidrule(lr){2-5}\cmidrule(l){6-7}
&
\sig{key} (IL) &
\sig{exec_done} (DoS) &
\sig{gpio_dir} (UA) &
\sig{pc} (FI) &
Precision &
Recall \\
\midrule
PDG-only
  & 20/42/0 & 10/443/0 & 3/6/0 & 23/430/0
  & 0.057 & 1.000 \\
Functional-only
  & 20/42/0 & 10/55/0 & 3/6/0 & 23/127/0
  & 0.196 & 1.000 \\
RTL-only
  & 18/18/2 & 7/11/3 & 3/6/0 & 14/20/9
  & 0.433 & 0.750 \\
Attack-only
  & 19/16/1 & 7/8/3 & 3/4/0 & 16/13/7
  & 0.523 & 0.804 \\
Feature-only
  & 19/3/1 & 8/5/2 & 3/1/0 & 18/8/5
  & 0.738 & 0.857 \\
Full SecID
  & 20/0/0 & 9/2/1 & 3/0/0 & 19/0/4
  & \textbf{0.962} & \textbf{0.911} \\
\bottomrule
\end{tabular}%
}
\vspace{-2em}
\end{table}

\subsubsection{Functional-Influence Threshold Sensitivity}
Here, we evaluate the Agent~2 threshold using the same four case studies. As shown in Table~\ref{tab:threshold_sensitivity}, both $\tau=0.025$ and $\tau=0.05$ retain all reference assets. However, lowering the threshold to 0.025 retains 345 candidates across the four cases, compared with 286 at 0.05, corresponding to 20.6\% more candidates for downstream analysis, which increases the computational cost. In contrast, increasing the threshold to 0.10 omits 13 reference assets from subsequent stages, increasing the risk of false negatives in the final SecID outputs. These results support the nominal choice of $\tau=0.05$ for our experiments.

\begin{table}[t]
\centering
\caption{Functional-influence threshold sensitivity}
\vspace{-2mm}
\label{tab:threshold_sensitivity}
\scriptsize
\setlength{\tabcolsep}{3pt}
\renewcommand{\arraystretch}{0.95}

\begin{tabular}{@{}llcccc@{}}
\toprule
\multirow{2}{*}{\textbf{Design}} &
\multirow{2}{*}{\textbf{Primary asset}} &
\multirow{2}{*}{\textbf{Ref.}} &
\multicolumn{3}{c}{\textbf{Retained}} \\
\cmidrule(lr){4-6}
& & &
$\boldsymbol{\tau=0.025}$ &
$\boldsymbol{\tau=0.05}$ &
$\boldsymbol{\tau=0.10}$ \\
\midrule
AES core    & key (IL)          & 20 & 20 & 20 & 20 \\
OpenMSP430  & exec\_done (DoS) & 10 & 10 & 10 & 7  \\
APB GPIO    & gpio\_dir (UA)    & 3  & 3  & 3  & 3  \\
OpenMSP430  & pc (FI)           & 23 & 23 & 23 & 13 \\
\bottomrule
\end{tabular}
\vspace{-2.0em}
\end{table}

\subsubsection{Formal Corroboration}\label{sec:formal}

We further use formal analysis to corroborate whether selected secondary assets participate in the modeled attack mechanism beyond bounded threat-feature experiments. This strengthens confidence in the secondary assets identified by SecID.

\textbf{Information Leakage.}
For openMSP430 \sig{r4}, JasperGold SPV models information leakage toward the UOP \sig{dbg_uart_txd}. With all SecID assets included, the baseline leakage witness is found in a median 4.39~s with a 573-cycle CEX trace across five runs. The assets are then restricted individually to assess their mediation in the leakage path. Restricting \sig{dbg_0.mem_data_rd}, in particular, increases the median SPV runtime to 80.63~s (18.36$\times$), while restricting \sig{dbg_0.mem_data} increases it to 49.09~s (11.18$\times$). Similar increases are observed for \sig{dbg_0.dbg_uart_0.xfer_buf_nxt} (7.17$\times$) and \sig{dbg_reg_din} (5.71$\times$). In all cases, SPV still finds an alternative leakage witness, indicating that these assets are not individually necessary but strongly mediate the modeled leakage path.

\textbf{Unauthorized Access.}
For NEORV32 \sig{lsu_priv}, JasperGold FPV checks the effect of excluding selected UA assets from the modeled bypass path. Within the evaluated property and control context, excluding either \sig{prv_level} or \sig{mstatus_mpp} leaves no bypass witness, whereas excluding \sig{mstatus_mprv} still leaves alternate bypass path(s). Despite contributing to the effective load/store privilege under MPRV operation, it is not a mandatory cutpoint for every modeled bypass route.

\textbf{Fault Injection.}
For CV32E40P, the 7-signal SecID cohort is compared in JasperGold FPV with equal-size randomly selected structurally connected sets across five seeds. All 10 evaluated SecID primary-impact checks produce a primary-asset fault witness, compared with a 17.9\% mean for the random cohorts. Moreover, independent attack-relevant corruption is observed in 11 of 14 SecID fault-path opportunities (78.6\%), compared with a 17.9\% random-cohort mean. These results show that the SecID-selected signals are enriched in fault sites capable of corrupting the primary asset.

\subsection{Final-Stage Repeatability}
We evaluate the repeatability of Agent~5 by executing it independently five times on four case studies while keeping the inputs, prompt, and model configuration fixed. For each case, the five resulting asset sets produce ten unique pairwise comparisons, with agreement measured using Jaccard similarity,
$J(A,B)=|A\cap B|/|A\cup B|$. As reported in Table~\ref{tab:repeatability}, the mean pairwise similarity ranges from 0.806 to 0.867, indicating generally consistent selections across repeated LLM runs.

\begin{table}[t]
\caption{Repeatability over 5 Final-stage LLM runs}
\vspace{-2mm}
\label{tab:repeatability}
\centering
\scriptsize
\renewcommand{\arraystretch}{1.05}
\setlength{\tabcolsep}{2.2pt}
\begin{tabular}{@{}llccc@{}}
\toprule
Design & Primary asset & Mean $J$ & Std. dev. & Range \\
\midrule
AES core
& \sig{key} (IL)
& 0.8064 & 0.0513 & 0.7083--0.8696 \\

OpenMSP430
& \sig{exec_done} (DoS)
& 0.8629 & 0.0544 & 0.7500--0.9167 \\

APB GPIO
& \sig{gpio_dir} (UA)
& 0.8667 & 0.1633 & 0.6667--1.0000 \\

OpenMSP430
& \sig{pc} (FI)
& 0.8354 & 0.0493 & 0.7619--0.9048 \\
\bottomrule
\end{tabular}
\vspace{-2.0em}
\end{table}

\section{Conclusion}
\label{sec:conclusion}
This paper introduces SecID, a threat-aware methodology that integrates static EDA analysis, bounded dynamic simulation, and LLM-based reasoning to identify security-critical secondary assets in hardware IPs and SoCs. Beyond identifying secondary assets, SecID provides threat-specific rationale through contextual attack mechanisms and quantitative threat-feature scores. This threat-aware characterization helps engineers understand how internal signals may contribute to different security compromises and supports the use of identified assets in subsequent security-verification tasks. By integrating threat-specific analysis into the secondary-asset identification flow, SecID helps systematize existing IP and SoC development processes while strengthening the broader hardware security verification landscape.

\section*{Acknowledgment}
This research was funded by the U.S. National Science Foundation (NSF) under the Faculty Early Career Development (CAREER) Program, Grant No.~2339971.

\clearpage
\bibliographystyle{IEEEtran}
\bibliography{references}

\end{document}